# Near-surface Azimuthal Magnetic Fields and Solar Activity Cycles

A.G. Tlatov

*Kislovodsk Mountain Astronomical Station of the Pulkovo observatory, Kislovodsk, Gagarina str. 100, 357700, Russia.*

**Abstract**

Variations of the azimuthal magnetic fields of the Sun in the 23-25 activity cycles of the activity cycles are considered. To identify azimuthal magnetic fields, the analysis of daily observations of LOS magnetic fields from the regions near the solar limb was performed. It is shown that with a sufficiently large averaging of the data, large-scale structures are distinguished that can be interpreted by horizontal magnetic fields directed along the East-West line. Azimuthal magnetic fields are visible both in the low-latitude zone and at high latitudes. Azimuthal fields at the same latitudes have opposite directions in the northern and southern hemispheres, and also change sign in even and odd cycles of activity.

The mechanism of formation of global azimuthal magnetic fields and their role in the cycle of solar activity is discussed. The near-surface azimuthal magnetic field is closely related to the activity cycle. Apparently, the azimuthal field is formed from U-shaped flux tubes of active regions (AR). Due to the presence of the tilt angle AR during differential rotation, the subsurface magnetic fields are pulled in the azimuthal direction. The role of azimuthal magnetic fields in solar activity cycles is considered. A scheme for generating a magnetic field according to a scheme different from Babcock- Leighton, dynamo models is proposed.

**Introduction**

Solar cyclicity manifests itself as an ordered process of excitation and dissipation of the magnetic field. There are general patterns in this process that manifest themselves in different cycles and at different latitudes. For example, Hale's law of polarity of sunspots, Maunder's law, regular reversal of the magnetic field in polar regions. These patterns are valid for both hemispheres of the Sun and can be traced throughout many cycles of activity. Such an organization of magnetic activity involves the formation of large-scale or global magnetic fields.

Solar activity is a consequence of the inductive action of fluid flows on large-scale magnetic fields (Charbonneau, 2020). The solar photospheric magnetic flux is formed by the rise of toroidal magnetic fields from the depths of the Sun in the form of active regions (AR), mainly in the form of bipolar structures. The total magnetic flux of each individual AO is approximately zero. But observations show (Babcock, 1964) that with the development of the activity cycle at high latitudes, unipolar large-scale magnetic fields of different polarities are formed in different hemispheres. The global magnetic field is most well manifested in the minima of solar activity, when the large-scale magnetic field takes configurations close to the dipole.

The polar fields of the Sun, located at the heliographic poles of the Sun, have a large-scale unipolar distribution covering a range of latitudes from about $\pm 50^{\circ}$-$60^{\circ}$ during most phases of the solar cycle, except for the period they change magnetic polarity. It is believed that their magnetic configuration is relatively simple, with predominantly almost vertical lines of force (Petrie 2023).

But the question of the existence of horizontal magnetic fields remains open. (Tsuneta et al., 2008; Petrie 2020; ). This is partly due to the difficulties of observing horizontal magnetic fields on the Sun. There are local horizontal magnetic fields associated with local sources, such as magnetic loops (Yabar et al, 2018). But to date, there is insufficient evidence of the existence of long-lived large-scale horizontal fields. Such fields can play a particularly important role in high-latitude regions, where they affect the formation of the coronal magnetic field on a large spatial scale, as well as because of their role in the cycle of solar activity (Petrie, 2015).

In this paper, the observations of LOS magnetograms are analyzed, but the main attention is paid to the allocation of horizontal large-scale magnetic fields. Based on the observational data, a mechanism for the formation of a near-surface magnetic field and the formation of sources for generating the next cycle of activity is proposed.

**Identification of azimuthal magnetic fields**

Direct measurements of horizontal magnetic fields can be performed by observing the full magnetic field vector on the photosphere. At the same time, there are difficulties in registering and interpreting the signal of the transverse component of the field (Petrie 2023). The sensitivity of magnetographs to transverse fields with the Zeeman effect is usually an order of magnitude lower than to longitudinal signals. The Zeeman effect for linear polarization is proportional to the square of the transverse field strength and is therefore weaker than the Zeeman effect for circular polarization, which has a linear dependence on the longitudinal field strength. Because of this quadratic dependence, the orientation of the transverse field suffers from the so-called 180-degree azimuth ambiguity, which must be resolved.

Since the observation of horizontal magnetic fields is difficult, we will use the observation data of the longitudinal component of the magnetic field. We will consider areas near the solar limb where the azimuthal component of the magnetic field can be seen at a small angle. To better determine the azimuthal component, we can consider the difference in the intensity of magnetic fields near the eastern (E) and western limbs (W). The main data in this work were the observation data of magnetic fields HMI/SDO. To reduce noise, in the period 2010-2023, we processed five images for each day at times close to 00:00, 05:00, 10:00, 15:00, and 20:00 UT. The regions separated from the central meridian φ by φ= $80^o$ -$89^o$ in longitude and $5^o$ wide in latitude near the E and W limbs were measured. For each day of observations, by averaging 5 observations at different points in time, we calculated the average values separately for the eastern BE and western BW limbs. To plot the graphs, we used values proportional to the square root of the intensity of the magnetic field, taking into account the sign of the magnetic field. This made it possible to improve the contrast on the diagrams for fields of different intensities.

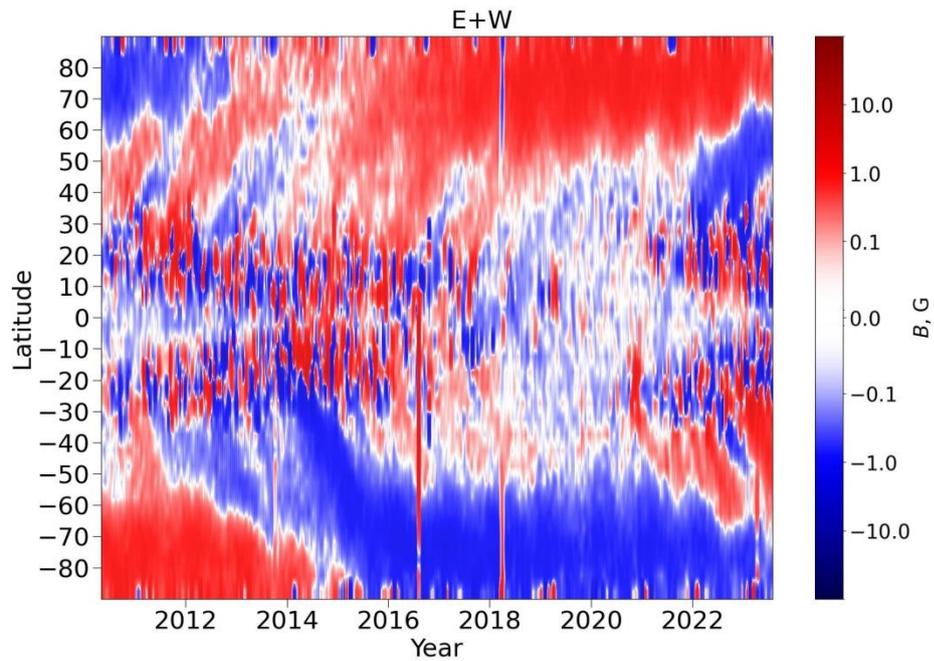

**Fig. 1**. Measurements of the magnetic field near the limb of the Sun according to SDO/ HMI data. The sum of the magnetic field intensities on the eastern and western limb is presented ($B_E+B_W$).

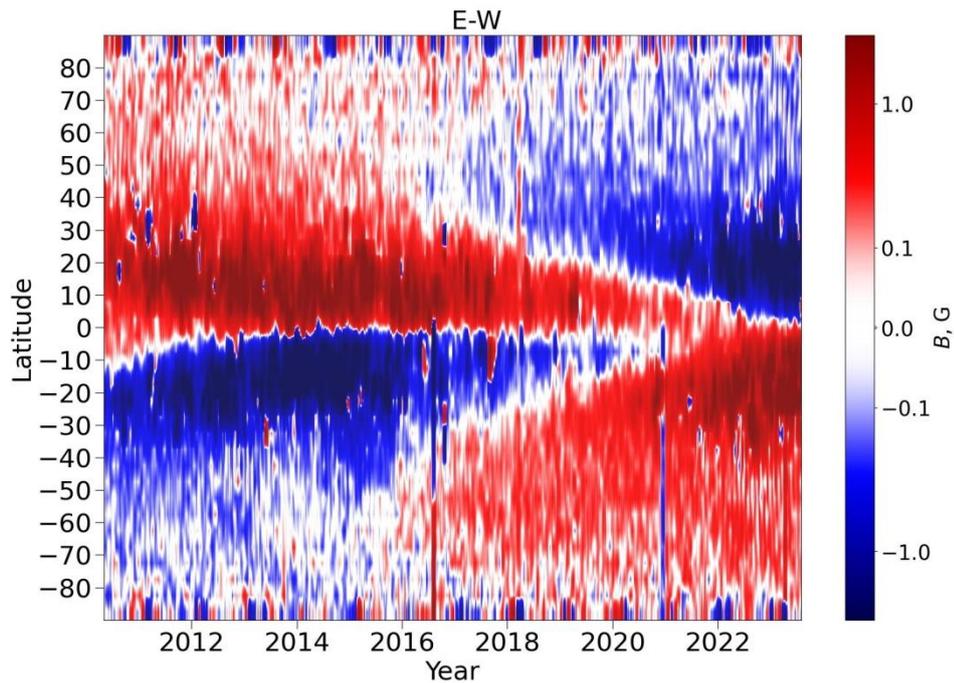

**Fig. 2**. The difference in magnetic field measurements at the eastern and western limb ($B_E-B_W$) according to SDO/ HMI data.

Figure 1 shows the distribution in latitude-time coordinates for the sum of the values of $B_E+B_W$, more precisely for ($B_E+B_W$). The distribution is close to the distribution for magnetic fields when sampling near the central meridian, although with different amplitudes of magnetic fields, since strong magnetic fields, for example in the sunspots umbra, the magnetic field is directed mainly vertically. The region of existence of AR in the form of "butterflies" and unipolar regions at high latitudes are visible. This diagram shows that the formation of polar regions occurs during the drift of fields with a polarity corresponding to the magnetic polarity of the tail sunspots in the form of flux pulses.

Figure 2 shows a diagram of the magnetic field difference ($B_E-B_W$). To reduce noise, daily values at each latitude interval were smoothed by a sliding window of 27 days. The distribution on Figure 2 in comparison with Figure 1 has changed significantly. We see large areas of fields of the same sign changing with the solar cycle. The distribution is asymmetric in the hemispheres. In the region of the formation of sunspots, we see unipolar regions. Latitudinal drift of regions of the same sign is observed during the activity cycle from the middle latitudes to the equator.

But we also see an ordered structure of the assumed azimuthal magnetic field at high latitudes. Comparing Fig. 1 and Fig. 2 we see that azimuthal magnetic fields of the N25 cycle appeared at latitudes ~50-60$^o$ in 2015-2016, that is, several years before the appearance of sunspots of the 25th cycle. Azimuthal fields of the 24th cycle near the equator disappear 2020-2022. We can assume that the duration of the existence of azimuthal fields in the cycle of activity is ~ 16-18 years, which is similar to the hypothesis of an extended cycle of activity.

At high latitudes, azimuthal magnetic fields drift to the poles, which may be due to a transfer with meridional circulation. At the same time, we do not observe any drift of the flux pulses from the sunspot zone to the poles, as in Fig. 1. It is possible that azimuthal magnetic fields have a significantly longer formation time and are more resistant to solar activity pulses.

Figures 3, 4 show separately the values of $B_E$ and $B_W$, respectively. At latitudes above 50$^o$, both on the eastern limb and on the western limb, we observe regions of the same polarity. The sign of the magnetic field $B_E$ and $B_W$ coincides with the sign of the magnetic fields for sunspots of tail polarity in activity cycles. At latitudes below 30$^o$, the magnetic field polarities on the eastern and western limbs (Fig. 3,4) have the opposite sign. Perhaps this is due to the Wilson effect.

The structure of the polarity distribution of magnetic fields near the limb is also confirmed by the data of other magnetographs. For these purposes, we also used the observation data of the SOHO/ MDI magnetograph in the period 1996-2011. As well as for the analysis of SDO/ HMI data, we used several observations per day. To measure the intensity of the magnetic field, the longitude interval $\varphi= 45^o -80^o$ was chosen. The latitude interval for averaging has not changed and is 5$^o$. Figure 5 shows the intensity difference on the eastern and western limbs ($B_E-B_W$). The data was smoothed by a sliding window 54 days wide.

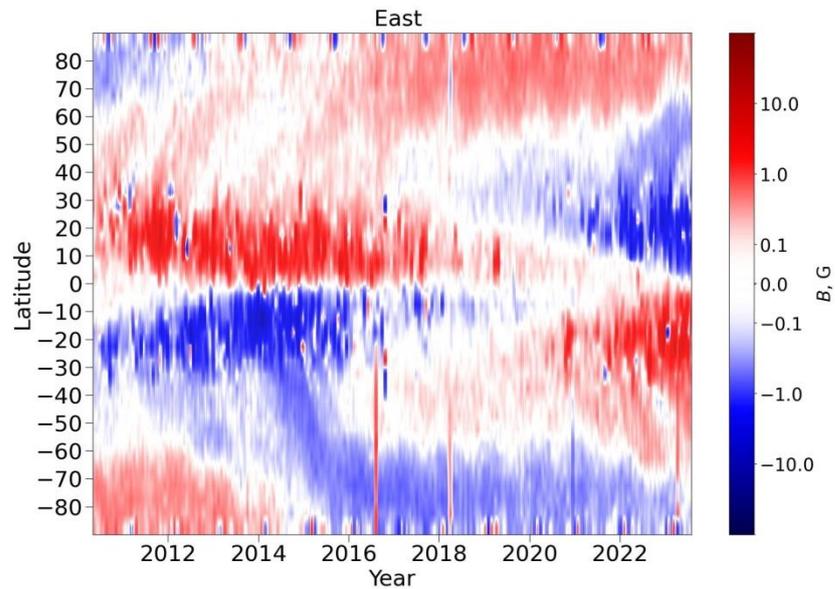

**Fig. 3.** Diagram of the distribution of the magnetic field near the eastern limb $B_E$ according to SDO/ HMI data.

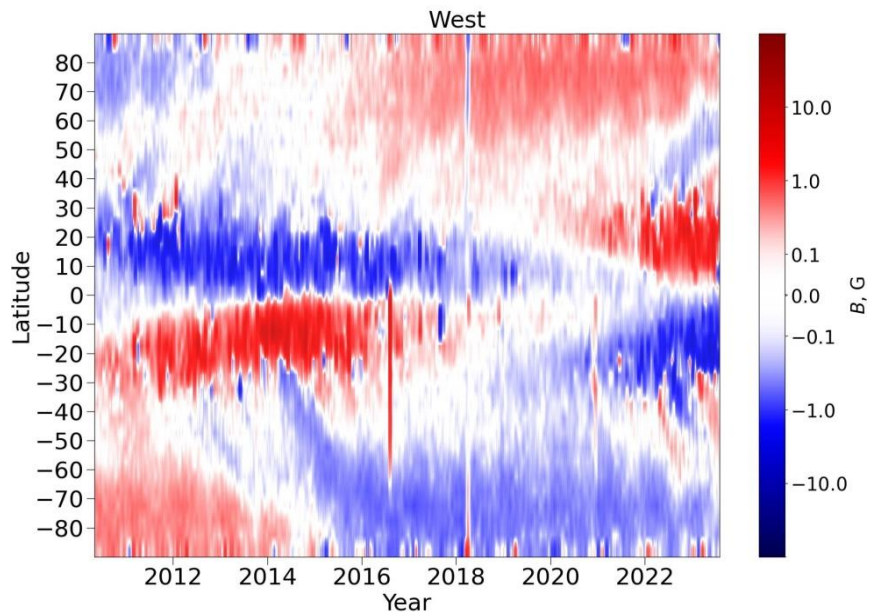

**Fig. 4.** Diagram of the distribution of the intensity of the magnetic field near the western limb of $B_W$ according to SDO/ HMI data.

After 2004, strong annual variations are visible, probably related to the noise component in the observations. But in general, there is a structure similar to the structure in Figure 2, taking into account the change of polarity signs in the 23rd cycle.

Figure 6 shows the BE and BW diagrams according to the SOHO/MDI magnetograph data. As well as for the SDO/HMI data in the low-latitude zone, the polarity of $B_E$ and $B_W$ coincides with the polarity sign of the tail and leading parts of the sunspots, respectively. At high latitudes, as well as in Fig. 3, 4 the polarity of the magnetic fields on the eastern and western limbs coincide with each other.

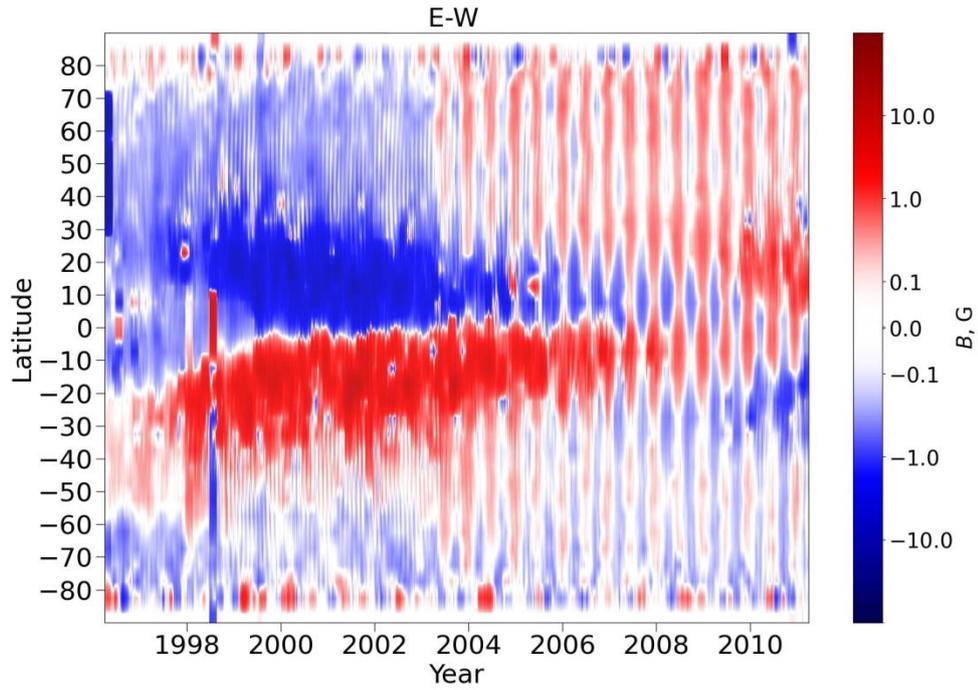

**Fig. 5**. The difference in magnetic field measurements at the eastern and western limb ($B_E$-$B_W$) according to SOHO/ MDI data.

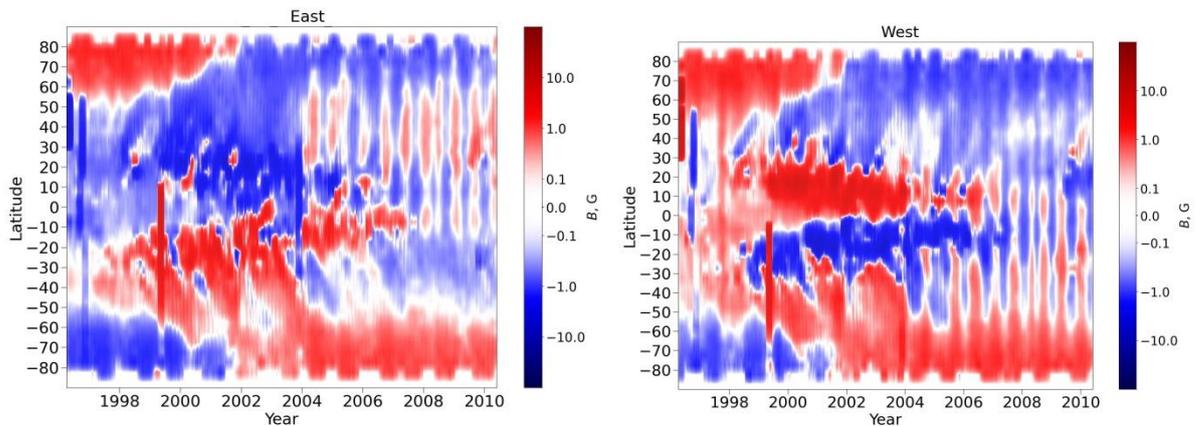

**Fig. 6**. Diagram of the magnetic field intensity distribution near the eastern limb $B_E$ (left) and the western limb $B_W$ (right) according to SOHO/ MDI data.

**Formation of a near-surface magnetic field**

From the analysis of daily observations of the longitudinal magnetic fields of the full disk of the Sun, we found a difference in the directions of the magnetic fields at the eastern and western limbs of the Sun. The time-latitude diagram (Figures 2, 5) shows that the regions of the magnetic field intensity difference parameter form long-lived large-scale structures, both in the low-latitude and high-latitude regions of the Sun. Perhaps this indicates the existence of a large-scale azimuthal magnetic field. Large-scale azimuthal magnetic fields can significantly change the generally accepted pattern of solar cycling generation.

Currently, one of the most popular models is the Babcock-Leighton dynamo (Babckock, 1959; Leighton, 1964). In this model, the toroidal magnetic field from the generation zone floats up as active regions to the surface of the Sun. Further, the magnetic fields of the tail parts of the active regions drift to the poles, and the magnetic fields of the parts of the leading polarities of different hemispheres are mutually annihilated through the equator. As a result, a new poloidal magnetic field of the dipole type is created, which sinks to the generation zone and, together with differential rotation, forms a toroidal magnetic field of a new cycle.

The occurrence of the AO flow occurs in cycles with an average period of ≈ 11 years at low latitudes (latitude ± 40°). Further, the magnetic flux is dispersed over the solar surface due to the diffusion effect of convective solar supergranulation flows, as well as differential rotation and meridional circulation (Wang, Nash, and Sheeley, 1989; Charbonneau, 2020).

At the same time, it can be seen from observations that the change of the sign of a large-scale magnetic field begins almost immediately after the appearance of a new AR cycle (Fig. 1). Sunspots at this moment are located at sufficiently high latitudes ~ $30^o$ far from the equator. Therefore, it is not entirely clear how the magnetic fields of the leading polarity of AR can be mutually destroyed through the equator with the fields of the other hemisphere. Also, the existing meridional circulation prevents the diffusion displacement of magnetic fields in the direction of the opposite hemisphere. Thus, the mechanism of formation of a large-scale magnetic field remains not fully understood.

The patterns of the occurrence of flows are the key to the cyclic interaction of AR and polar fields. According to Hale's law of polarity, bipoles in the northern or southern hemisphere of the Sun during a given cycle of activity have a leading/driven magnetic flux of the same polarity, with opposite polarity in the northern and southern hemispheres. This polarity is reversed in each next cycle. Moreover, the flux of the leading polarity of magnetic bipoles is on average located at a lower latitude than the flux of the tail polarity. This bias is called Joy's law. (Hale et al., 1919). Thus, the bipoles appear to be tilted relative to the equator. Thus, the bipoles appear to be tilted relative to the equator. Joy's law leads to the fact that the trailing polarity of the bipolar AR dominates at high latitudes.

Probably, the assumption about the dissipation of magnetic fields through the equator is wrong. We can assume that the multiple rising of AO in each hemisphere is important for solar cyclicity, as well as the shift of the average latitude of AO to the equator, called Maunder butterflies and Joy's law. Consider the evolution of active regions.

Suppose that the flux tubes of sunspots do not extend far into the convective zone, but form a U-shaped configuration. Since the trailing parts of the AO lie on average higher in latitude than the leading parts, the U-configuration power tube is affected by differential rotation. At the same time, the distance between the geometric centers of the trailing and leading polarities will increase over time. That is, the U-tube will elongate in the azimuthal direction.

Consider the evolution of two neighboring active regions AR1 and AR2. Due to Maunder's law, the latitude of the area that surfaced later in time (AR2) is on average closer to the equator. And due to Jay's law, the latitudinal distance between the leading sunspot of the first AR1 (L1) and the trailing sunspot of the second AR2 (T2) is, on average, less than between the tailing sunspot of the first AR1 (T1) and the leading sunspot of the second AR2 (L2). This leads to a preferential reconnection of the magnetic fields L1 and T2 and the formation of a power tube between T1 and L2 (Figure 7). By this mechanism, an azimuthal magnetic field is formed under the photosphere.

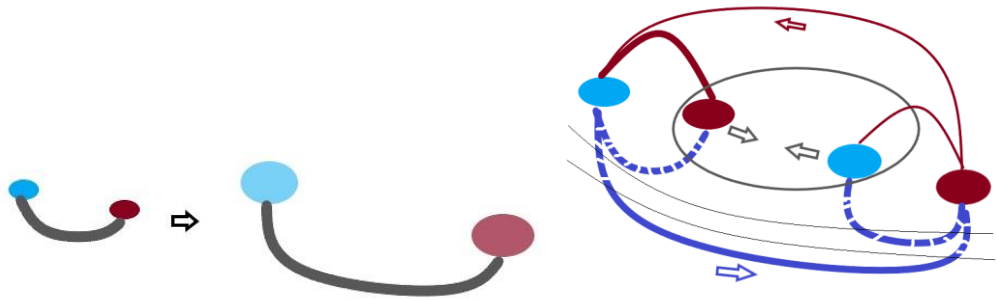

**Fig. 7**. Diagram of the formation of an azimuthal magnetic field under the photosphere. a) Stretching of the magnetic field tube AR in the azimuthal direction as a result of differential rotation. b) The interaction of the leading and tail parts of the two AR and the formation of a new azimuthal field tube..

Azimuthal magnetic fields are formed under the photosphere by the ascent and interaction of many AO regions in each hemisphere. Figure 8 shows a diagram of the formation of a near-surface azimuthal magnetic field. U-shaped tubes form an azimuthal field, with the direction opposite to the toroidal magnetic field in the generation zone.

Due to the meridional transfer, the new field is moved to the generation zone (Figure 9). In addition to the azimuthal component, due to the geometry of the AR, a poloidal magnetic field is also formed (see Figure 8) This poloidal magnetic field forms the observed distribution of a large-scale magnetic field at the poles. Also, the poloidal component, when it is transferred to the generation zone, provides strengthening of the toroidal component at the base of the convective zone.

The presented scheme of magnetic field generation is based on the observed effects, such as Maunder's law, Joy's law, meridional circulation and differential rotation. This scheme does not require mutual destruction of the fields of the leading regions of the northern and southern hemispheres across the equator. But there are two zones of formation of toroidal magnetic fields at the base of the convective zone and under the photosphere. It is possible to modify this scheme and exclude the conveyor transfer of the magnetic field to the generation zone to the base of the convective zone with meridional circulation. In this case, the new toroidal field can be immersed from the photosphere to the generation zone by other mechanisms, for example, by flows of matter during supergranulation.

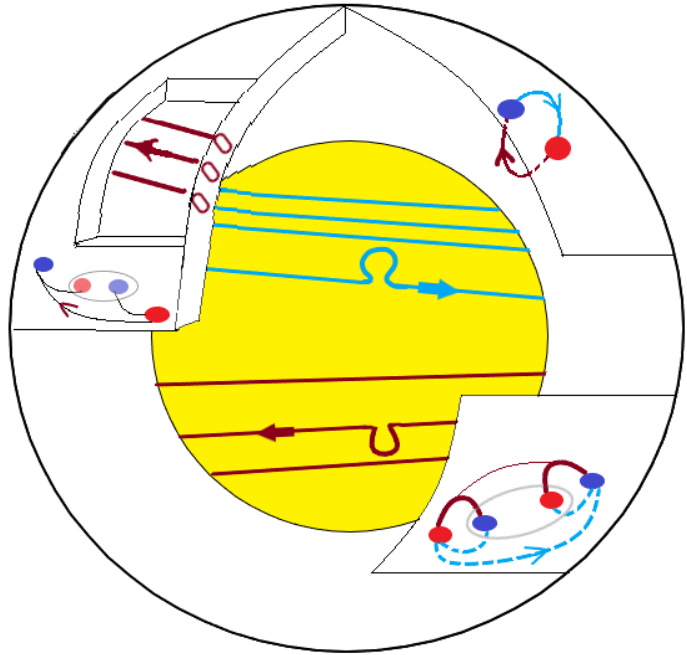

**Fig. 8.** Scheme of existence of toroidal fields on the Sun.

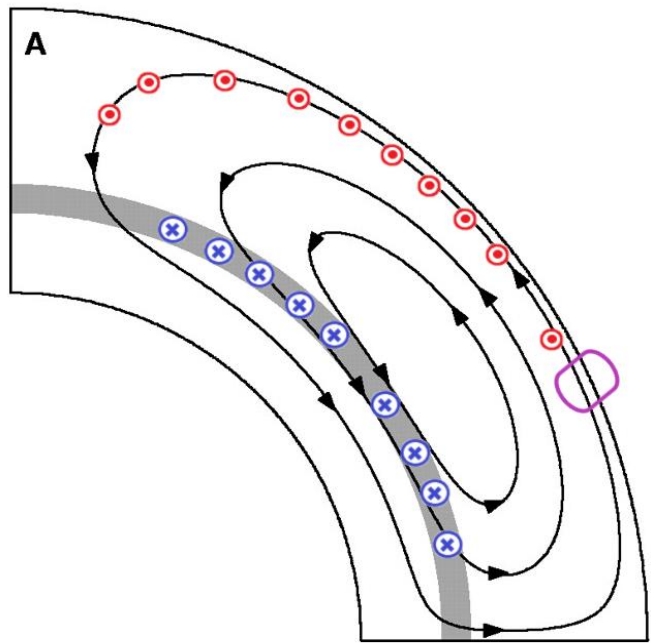

**Fig. 9**. A possible scheme for the transfer of a new azimuthal field by a meridional circulation conveyor (black lines with arrows). The blue color shows the toroidal magnetic fields that lead to the appearance of AR (lilac loop) in the current cycle. The red color shows the subsurface azimuthal fields that appeared as a result of the ascent of the AR.

## Conclusion

In this article, the analysis of daily magnetograms according to the data of SDO/ HMI and SOHO/ MDI magnetographs is carried out in order to isolate large-scale horizontal magnetic fields. LOS magnetograms were used for this, and measurements were carried out near the solar limb in the eastern and western hemispheres of the Sun. To distinguish the signal above the noise, several observations per day and time averaging were used. This made it possible to identify long-lived structures that can be interpreted by the existence of a large-scale azimuthal magnetic field $B^P_T$.

     A scheme for the formation of a near-surface azimuthal field $B^P_T$ is proposed. Such a magnetic field is formed when the U-shaped tubes of the surfaced active regions are pulled out. The set of surfaced AR forms a single axisymmetric magnetic field having both azimuthal and poloidal components. In favor of this, a quieter change in the azimuthal field (Figure 2) can serve, compared with a more sporadic change in individual components (Figures 3, 4). Due to the presence of tilt angle AR, differential rotation at the surface stretches the U-structures, strengthening the azimuthal field at the surface.

     During the cycle, the new $B^P_T$ field is transferred to the poles, where the poloidal part is visible as unipolar high-latitude caps of a large-scale magnetic field. The new field is transferred to the generation zone. The transfer can be carried out by conveyor transport by meridional circulation or by other mechanisms of dipping of this field.

     The near-surface azimuthal field of the component has the opposite polarity in different hemispheres and changes in sign in successive cycles of activity (Figure 2). The azimuthal field of the current cycle appears at mid-latitudes several years before the start of the sunspot cycle. It is possible that the source of the azimuthal field during this period are ephemeral regions.